\pdfoutput=1
\documentclass[aps,twocolumn,prb,showpacs,floatfix]{revtex4}
\usepackage{graphicx}
\usepackage{amsmath}
\usepackage{amssymb}
\usepackage{times}
\begin{document}

\title{Magnetic molecules created by hydrogenation of
Polycyclic Aromatic Hydrocarbons}
\author{J. A. Verg\'es}
\affiliation{Departamento de Teor\'{\i}a de la Materia Condensada,
Instituto de Ciencia de Materiales de Madrid (CSIC), Cantoblanco, 28049 Madrid,
Spain.}
\email{jav@icmm.csic.es}
\author{G. Chiappe}
\author{E. Louis}
\affiliation{Departamento de F\'{\i}sica Aplicada, Unidad Asociada del
CSIC and
Instituto Universitario de Materiales, Universidad
de Alicante, San Vicente del Raspeig, 03690 Alicante, Spain.}
\author{L. Pastor-Abia}
\author{E. SanFabi\'an}
\affiliation{Departamento de Qu\'{\i}mica F\'{\i}sica, Unidad Asociada del
CSIC and
Instituto Universitario de Materiales, Universidad
de Alicante, San Vicente del Raspeig, 03690 Alicante, Spain.}
\date{July 30, 2008}
\begin{abstract}
Present routes to produce magnetic organic-based materials adopt a
common strategy: the use of magnetic species (atoms, polyradicals, etc.)
as building blocks. We explore an alternative approach which
consists of selective hydrogenation of Polycyclic Aromatic Hydrocarbons.
Self-Consistent-Field (SCF) (Hartree--Fock and DFT) and multi-configurational
(CISD and MCSCF) calculations on coronene
and corannulene, both hexa-hydrogenated,
show that the formation of stable high spin species is possible.
The spin of the ground states is
discussed in terms of the Hund rule and Lieb's theorem for bipartite
lattices (alternant hydrocarbons in this case).
This proposal opens a new door to magnetism in the organic world.
\end{abstract}
\maketitle

\section{Introduction}
Two successful routes that are being actually followed to produce
magnetic organic materials \cite{Mi00} are the addition of magnetic atoms
\cite{JL07} and the use of polyradicals \cite{Ra05}. In particular,
carbon-based nickel compounds that show spontaneous field-dependent
magnetization and hysteresis at room temperature, have been recently
synthesized \cite{JL07}. Moreover, the combination of two radical
modules with different spins has allowed the obtaining of organic polymers with ferro-
or antiferromagnetic ordering \cite{RW01}. Research on molecules containing
polyradicals goes back to the early nineties \cite{Ra05,RR96,SM04,Kr05}
and has produced a variety of results as, for example, the
synthesis of high spin organic molecules.
In some of these molecules the failure of Hund's rule has been
demonstrated \cite{RR96}. On the other hand, experimental and theoretical
evidence has been recently presented  indicating that 5-dehydro-m-xylylene or DMX was the first example
of an organic tri-radical with an open-shell doublet
ground-state\cite{SM04,Kr05}.
Both methods share a common strategy: the use of ingredients
(either radicals or atoms) that provide a finite spin.

In this work we follow a different approach. Specifically, we predict
the existence of spin polarized organic molecules derived
from non magnetic $\pi$-conjugated Polycyclic
Aromatic Hydrocarbons (PAHs) by selective hydrogenation
of their peripheral C atoms. High hydrogenation of PAHs
has been proposed as a method for hydrogen storage \cite{PS06}.
More recently, the feasibility of double hydrogenation
of those compounds has been investigated theoretically \cite{ZC07}.

Our work is inspired upon Lieb's theorem for bipartite lattices that
shows the appearance of magnetism whenever they are unbalanced \cite{Li89}.
According to Lieb, if a nearest neighbor
model with a local on-site interaction is applicable to a bipartite lattice,
the spin multiplicity of the ground state is $|N_A - N_B| + 1$,
where $N_A$ and $N_B$ are the number of atoms in each sublattice.
Most PAHs are alternant hydrocarbons
where carbon atoms can be separated into two disjoint subsets so that an
atom in one set only has neighbors in the other set
(Figs. 1 and 2 show a colored version of the partition).
The same theorem has been used to support the existence of
magnetism in graphene ribbons and islands \cite{grafenos}.
All work we know is based on single-determinantal methods, i.e.,
on a more or less sophisticated form of Self-Consistent-Field (SCF)
calculation. Let us remark that being $\pi$-orbital magnetism a direct
result of the strong correlation among $\pi$-electrons, only methods
designed explicitly to catch these effects (like CISD and MCSCF, used
in our work) can help to resolve the doubts regarding the appearance of
magnetism in graphite-derived systems. 

\begin{figure}
\includegraphics[width=\columnwidth]{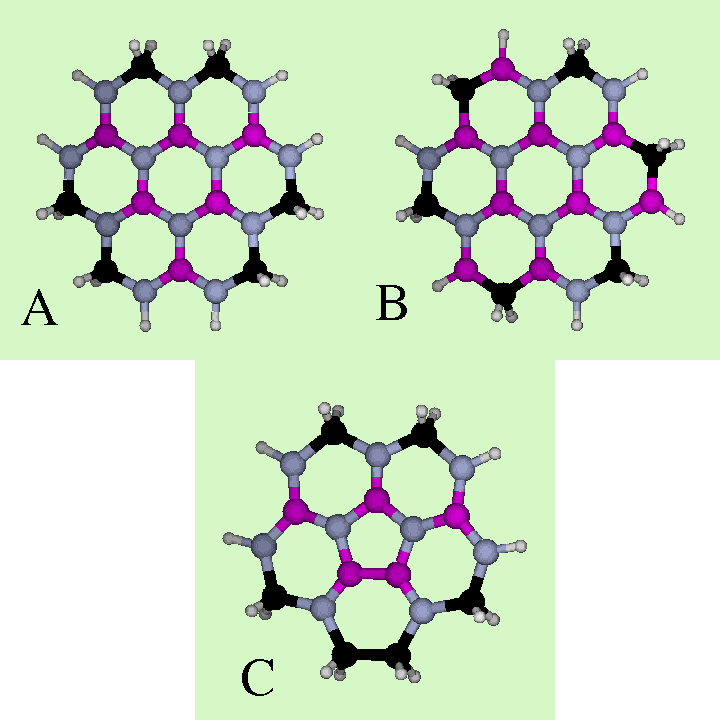}
\caption{(color online)
1,4,5,8,9,12-hexahydrocoronene (A, hereafter referred to as
$D_{3h}$ according to its symmetry group),
1,3,5,7,9,11-hexahydrocoronene (B, hereafter referred to as
$C_{3h}$) and planar 1,4,5,6,7,10-hexahydrocorannulene
(C, hereafter referred to as $C_{2v}$).
Saturated carbon atoms are represented by black
symbols while dark gray (magenta) and light gray symbols are used
to distinguish carbon atoms belonging to different sublattices.
Corannulene is a non-alternant hydrocarbon, that is, a frustrated cluster
of carbon atoms (note the fully magenta bond between two magenta atoms).}
\label{fig:moleculesI}
\end{figure}

The rest of the paper is organized as follows. The {\it ab initio} methods
(both mono- and multi-determinantal) used in this work are discussed in some
detail in section II, while the results obtained with those methods are
reported and discussed in section III. Section IV in turn is devoted to
the analysis of the {\it ab initio} results by means of model Hamiltonians,
in particular the Hubbard and the Pariser-Parr-Pople Hamiltonians. Finally,
the conclusions of our work are gathered in section V.

\begin{figure}
\includegraphics[width=\columnwidth]{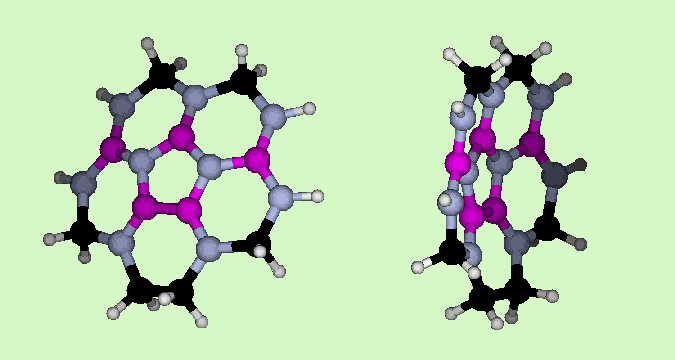}
\caption{(color online)
Two views of curved 1,4,5,6,7,10-hexahydrocorannulene
in the calculated stable geometry (hereafter referred to as $C_1$).
As in Fig. 1, black symbols indicate carbon atoms forming only
single bonds while dark gray (magenta) and light gray symbols
denote each of the two sublattices in which carbon atoms can be
separated.}
\label{fig:moleculesII}
\end{figure}

\begin{table}
\caption{
Total energies (in Hartrees) for atomic hydrogen,
molecular hydrogen, coronene C$_{24}$H$_{12}$,
corannulene  C$_{20}$H$_{10}$,
two molecules obtained from hexahydrogenation
of coronene and one derived from hexahydrogenation of corannulene
(in the latter case the results correspond to the planar geometry
shown in Fig. 1C).
The results were obtained using three basis sets
(MIDI, cc-pVDZ and cc-pVTZ),
two SCF methods (RHF and RB3LYP) and one
multi-configurational method CISD. The number of
occupied (m) and empty (n) $\pi$ Molecular Orbitals included
in the CISD calculations as well as the number of electrons that
fill them (N) is indicated as (m+n,N). Small stars emphasize
the spin multiplicity of the more stable state.}
\begin{tabular}{|c|c|ccc|}
\hline
MOLECULE &  METHOD     &    \multicolumn{3}{c|}{BASIS SET} \\
\cline{3-5}
        &             &  MIDI  & cc-pVDZ & cc-pVTZ   \\
\hline
\hline
H        & RHF        & -0.4970 & -0.4993 & -0.4998 \\
        & RB3LYP      & -0.4953 & -0.4979 & -0.4988 \\
\hline
H$_2$   & RHF         & -1.1217 & -1.1287 & -1.1330 \\
        & RB3LYP      & -1.1623 & -1.1668 & -1.1733 \\
\hline
Coronene &  RHF                   & -910.4869 & -916.0197 & -916.2293  \\
   C$_{24}$H$_{12}$   & RB3LYP    & -915.9341 & -921.3874 & -921.6253  \\
\hline
Corannulene &  RHF                & -758.6127 & -763.2326 & -763.4078  \\
   C$_{20}$H$_{10}$   & RB3LYP    & -763.1633 & -767.7138 & -767.9092  \\
\hline
\hline
C$_{24}$H$_{18}$ & RHF   & -913.6339 & -919.2239 & -919.4337 \\
$\star$ $D_{3h}$ (S=3) $\star$ &  RB3LYP & -919.1814 & -924.6701 & -924.9159 \\
        & CISD(11+3,16)  & -913.7112 & -919.2905 & -919.4939 \\
\hline
C$_{24}$H$_{18}$& RHF    & -913.3714 & -918.9826 & -919.2040  \\
$D_{3h}$ (S=0)  & RB3LYP & -919.0751 & -924.5639 & -924.8175  \\
        & CISD(8+6,16)   & -913.4988 & -919.0286 & -919.2431  \\
\hline
\hline
C$_{24}$H$_{18}$& RHF    & -913.5690 & -919.1638 & -919.3742 \\
$C_{3h}$ (S=3)  & RB3LYP & -919.1228 & -924.6146 & -924.8620 \\
        & CISD(11+3,16)  & -913.6153 & -919.2027 & -919.4078 \\
\hline
C$_{24}$H$_{18}$&  RHF   & -913.7732 & -919.3607 & -919.5746 \\
$\star$ $C_{3h}$ (S=0) $\star$  & RB3LYP & -919.3423 & -924.8230 & -925.0730 \\
        & CISD(8+6,16)   & -913.8433 & -919.4158 & -919.6110 \\
\hline
\hline
C$_{20}$H$_{16}$ & RHF   & -761.9008 & -766.5625 & -766.7408 \\
$\star$ $C_{2v}$ (S=2) $\star$   &RB3LYP & -766.5503 & -771.1250 & -771.3348 \\
        & CISD(8+4,12)   & -761.9652 & -766.6291 & -766.8014 \\
\hline
C$_{20}$H$_{16}$ & RHF   & -761.7591 & -766.4418 & -766.6266 \\
$C_{2v}   $ (S=0) &  RB3LYP & -766.4913 & -771.0760 & -771.2842 \\
        & CISD(6+6,12)   & -761.8264 & -766.4643 & -766.6588 \\
\hline
\end{tabular}
\label{table:energy-TE}
\end{table}

\section{{\it Ab initio} calculations: Methods and Numerical Procedures}
Calculations of the spin states of the molecules of Figs. 1 and 2 were done
using the following basis functions sets: MIDI \cite{HUZINAGA-1984}, cc-pVDZ
and cc-pVTZ \cite{JCP-90-1007}.
Although the latter set guarantees a sufficient precision,
varying the dimension of the variational space allowed to check the
reliability of our results. SCF calculations were carried out at the
Restricted-Hartree-Fock (RHF) level and by means of the hybrid
density functional RB3LYP \cite{JCP-98-5648,PRB-37-785,TCA-37-329}.
In both cases the Restricted-Open-Shell variant was used in order
to get well-defined total spin values \cite{UHF}.
In order to check the accuracy of
the description of the correlation energy of partially filled
$\pi$-shells, multi-configurational wave-functions calculations
were also performed. Configuration Interaction with Single
and Double excitations (CISD) calculations \cite{BS79} were carried out
in all cases, while some checks were also made by means of the
Multi-Configurational SCF (MCSCF) on the fully optimized set in the
active space version \cite{ARPC-49-233,ACP-69-399}.
The active space was generated within the following windows (m+n,N) of m
occupied and n empty $\pi$ Molecular Orbitals (MO) filled with N electrons:
hexahydrogenated coronene S=0, (8+6,16) and S=3, (11+3,16), and planar
hexahydrogenated corannulene S=0, (6+6,12) and S=2, (8+4,12).
Other $\pi$-MO lie
excessively far from the HOMO-LUMO gap to give a sizable contribution.
Geometries were only optimized at the SCF (RB3LYP) level.
The geometry of 6H-corannulene was
optimized for both its planar metastable form and its curved stable form
(see Figs. 1C and 2).
However, in order to allow a discussion in terms of $\pi$-orbital
models, the results for the energies of its spin states discussed
hereafter correspond to the planar geometry.
Anyhow, energy differences between the spin states of the two
allotropes are very small (fragmentation energies for
both planar and curved geometries are reported below).
All quantum chemistry calculations were done using the GAMESS
program \cite{GAMESS}.

\begin{figure}
\includegraphics[width=0.58\columnwidth]{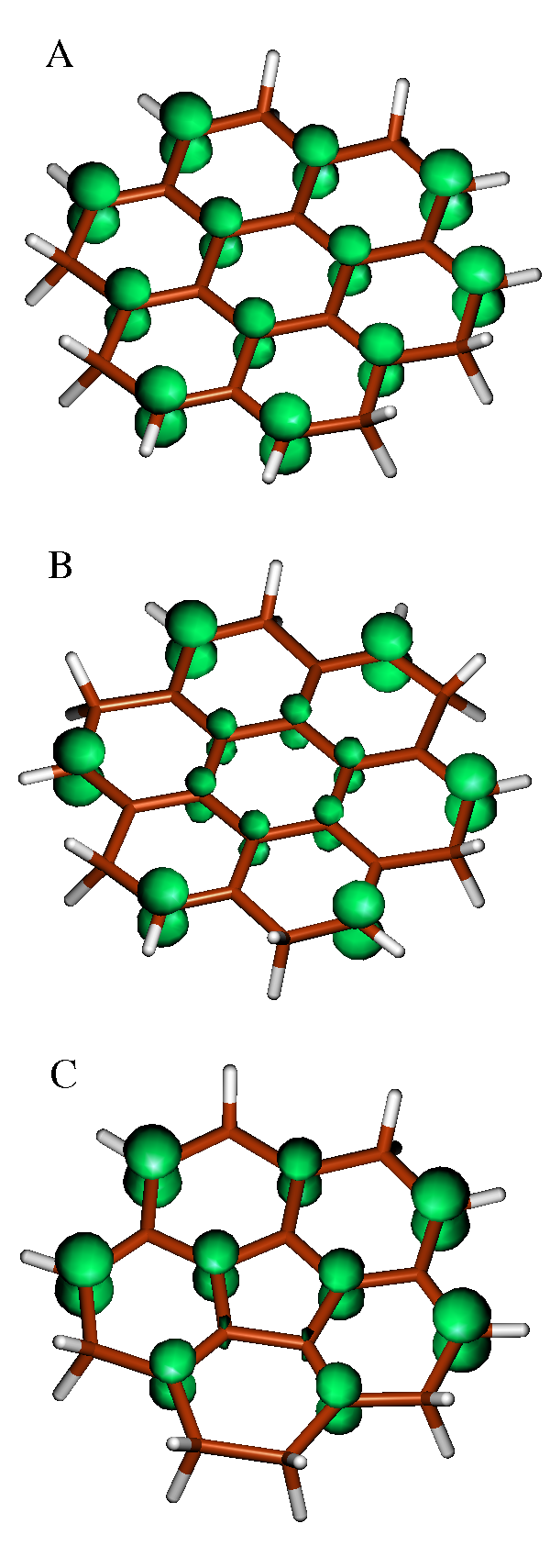}
\caption{(color online)
Total spin densities of 1,4,5,8,9,12-hexahydrocoronene and
1,3,5,7,9,11-hexahydrocoronene (both corresponding to septuplets, S=3)
and planar 1,4,5,6,7,10-hexahydrocorannulene (S=2 state).}
\label{fig:moleculesIII}
\end{figure}

\begin{table}
\caption{
Fragmentation energies (in Hartrees) of molecules derived from
hexahydrogenation of coronene and of corannulene. Total energy
differences are given both for atomic and molecular forms of
hydrogen. Data of Table I have been used and again small stars emphasize
the spin multiplicity of the more stable state.}
\begin{tabular}{|c|c|cc|cc|}
\cline{3-6}
 \multicolumn{2}{c|}{}  &  \multicolumn{2}{c|}{Atomic H} &  \multicolumn{2}{c|}
        {Molecular H$_2$}   \\
\hline
MOLECULE &  METHOD &  MIDI   & cc-pVTZ & MIDI   & cc-pVTZ \\
\hline
\hline
C$_{24}$H$_{18}$        & RHF     & -0.1651 & -0.2056 & 0.2181 & 0.1946 \\
$\star$ $D_{3h}$ (S=3) $\star$ & RB3LYP & -0.2755 & -0.2980 & 0.2396 & 0.2292 \\
\hline
C$_{24}$H$_{18}$        & RHF     &  0.0974 &  0.0242 & 0.4805 & 0.4243 \\
     $D_{3h}$ (S=0)     & RB3LYP  & -0.1692 & -0.1996 & 0.3459 & 0.3276 \\
\hline
\hline
C$_{24}$H$_{18}$        & RHF     & -0.1002 & -0.1460 & 0.2830 & 0.2542 \\
     $C_{3h}$ (S=3)     & RB3LYP  & -0.2170 & -0.2441 & 0.2982 & 0.2831 \\
\hline
C$_{24}$H$_{18}$        & RHF     & -0.3044 & -0.3465 & 0.0788 & 0.0537 \\
$\star$ $C_{3h}$ (S=0) $\star$ &RB3LYP  & -0.4365 & -0.4551 & 0.0787 & 0.0720 \\
\hline
\hline
C$_{20}$H$_{16}$        & RHF     & -0.3061 & -0.3342 & 0.0770 & 0.0660 \\
$\star$ $C_{2v}$ (S=2) $\star$ &RB3LYP  & -0.4152 & -0.4290 & 0.0999 & 0.0981 \\
\hline
C$_{20}$H$_{16}$        & RHF     & -0.1644 & -0.2200 & 0.2187 & 0.1802 \\
    $C_{2v}$ (S=0)      & RB3LYP  & -0.3562 & -0.3784 & 0.1589 & 0.1487 \\
\hline
\hline
C$_{20}$H$_{16}$        & RHF     & -0.3024 & -0.3259 & 0.0807 & 0.0743 \\
$\star$ $C_{1}$ (S=2) $\star$ & RB3LYP  & -0.4098 & -0.4198 & 0.1054 & 0.1073 \\
\hline
C$_{20}$H$_{16}$        & RHF     & -0.1607 & -0.2072 & 0.2224 & 0.1930 \\
    $C_{1}$ (S=0)       & RB3LYP  & -0.3471 & -0.3643 & 0.1680 & 0.1628 \\
\hline
\end{tabular}
\label{table:energy-md}
\end{table}

\section{{\it Ab initio} calculations: Results}
Total energies for the singlet and the relevant multiplet of
hydrogenated coronene $D_{3h}$, $C_{3h}$ and planar hydrogenated
corannulene $C_{2v}$ (A, B and C in Fig. 1) are reported in Table I.
It is first noted that whereas the energies obtained with the small
basis set MIDI and those obtained with the already large cc-pVDZ,
differ in 4-6 Hartrees (approximately 0.6\%), the
difference is reduced to 0.1-0.3 Hartrees (approximately 0.02\%)
when cc-pVDZ is replaced by the largest basis used in this work,
namely, the cc-pVTZ basis set. This indicates that convergence,
as far as the basis set is concerned, is rather acceptable.
In the case of hexahydrogenated coronene (briefly 6H-coronene),
results clearly show that,
no matter the method or the basis set used, the ground state of molecule
$D_{3h}$ is a septuplet and that of molecule $C_{3h}$ a singlet.
We have checked that other spin states lie between those two.
In molecule $D_{3h}$ the largest energy difference between the high spin
ground state and the singlet occurs for RHF (0.23-0.26 Hartrees).
This difference is reduced to approximately 0.1 Hartrees for RB3LYP,
increasing again using the CISD method.
On the other hand, all results for $C_{3h}$ conformation
show that the singlet is below the septuplet by more than 0.2 Hartrees.
Similar results are obtained for 6H-corannulene, although
energy differences are  slightly smaller.
Table I also reports total energy results for atomic and
molecular hydrogen, coronene and corannulene
that allow the calculation of fragmentation energies (Table II analysis).
These are negative relative to atomic hydrogen but not relative to the
molecular form. Therefore,
actual synthesis of the hydrogenated molecules would need
sophisticated reaction paths \cite{Albert}.
We also note that the singlet ground state of $C_{3h}$ hydrogenated coronene
is more stable than that of the molecule having a septuplet
ground state ($D_{3h}$). Presumably, other forms of 6H-corannulene
would also show deeper ground state energies than that of the studied magnetic
conformation. Note also that hydrogenation of the curved (stable) geometry
of corannulene (see Fig. 2) is 
slightly less favorable than that of its planar geometry
(compare results for $C_{2v}$ and $C_1$ in Table II).
Anyhow, as in the planar geometry, the quintuplet has a
lower energy than the singlet.

Fig. 3 depicts the total spin densities of the septuplet states (S=3) of
1,4,5,8,9,12-hexahydrocoronene and 1,3,5,7,9,11-hexahydrocoronene (A and B)
and the quintuplet (S=2) of planar 1,4,5,6,7,10-hexahydrocorannulene.
Concerning 1,4,5,8,9,12-hexahydrocoronene, the most appealing result is
that the spin density is finite only on the carbon atoms of one sublattice.
More precisely, spin density is located in the sublattice to which no
additional H atoms were attached.
This result is highly illustrative allowing some intuition on the reasons
for a magnetic ground state: electron-electron repulsion is minimized
because each electron avoids sitting at nearest-neighbors distances from the
others. However, in 1,3,5,7,9,11-hexahydrocoronene,
a molecule with a singlet ground state, the spin is equally spread over the
two sublattices implying larger electronic repulsions at
the central hexagon.
The case of 1,4,5,6,7,10-hexahydrocorannulene is even more interesting as,
being a frustrated molecule, at least one bond between atoms of the same
sublattice should be present. This is clearly visible in Fig. 3
once a sublattice is identified as the sites showing spin density while
the rest belong to the other sublattice (Colors in Fig. 1 have anticipated
this feature). We will show later that the model Hamiltonian calculations
for 1,4,5,6,7,10-hexahydrocorannulene show frustration
at the same bond than {\it ab initio} calculations (compare Figs. 3 and 4).
Having identified the atoms at each sublattice, it is tempting to use
the unbalance in the molecule ($N_A-N_B$=4) to predict the total
spin of the ground state using Lieb's formula. The result (S=2)
is in perfect agreement with numerical results.
This is particularly interesting as in principle Lieb's theorem
should only work on non-frustrated systems.

Spin multiplicity of the ground state of a molecule is usually predicted
by means of Hund rule applied to MO energies obtained by an
appropriate method. We have checked that the spin of the ground states
of the molecules here investigated is consistent
with the degeneracy of the HOMO that H\"uckel's method
gives for the skeleton of C atoms having an unsaturated $\pi$ orbital.
This is true not only for 6H-coronene, but also for
6H-corannulene. Although the extended H\"uckel's method used
by {\it ab initio} codes to initialize the self-consistency
process slightly lifts this degeneracy, the HOMO still appears as a narrow
bunch containing a number of orbitals compatible with the spin of the ground
states of the three planar molecules depicted in Fig. 1.
Then, as in Hund rule,
such a distribution of molecular orbitals favors high
spin ground states through a winning competition of interaction energy
gains against kinetic energy losses.


\begin{figure*}
\includegraphics[width=2\columnwidth]{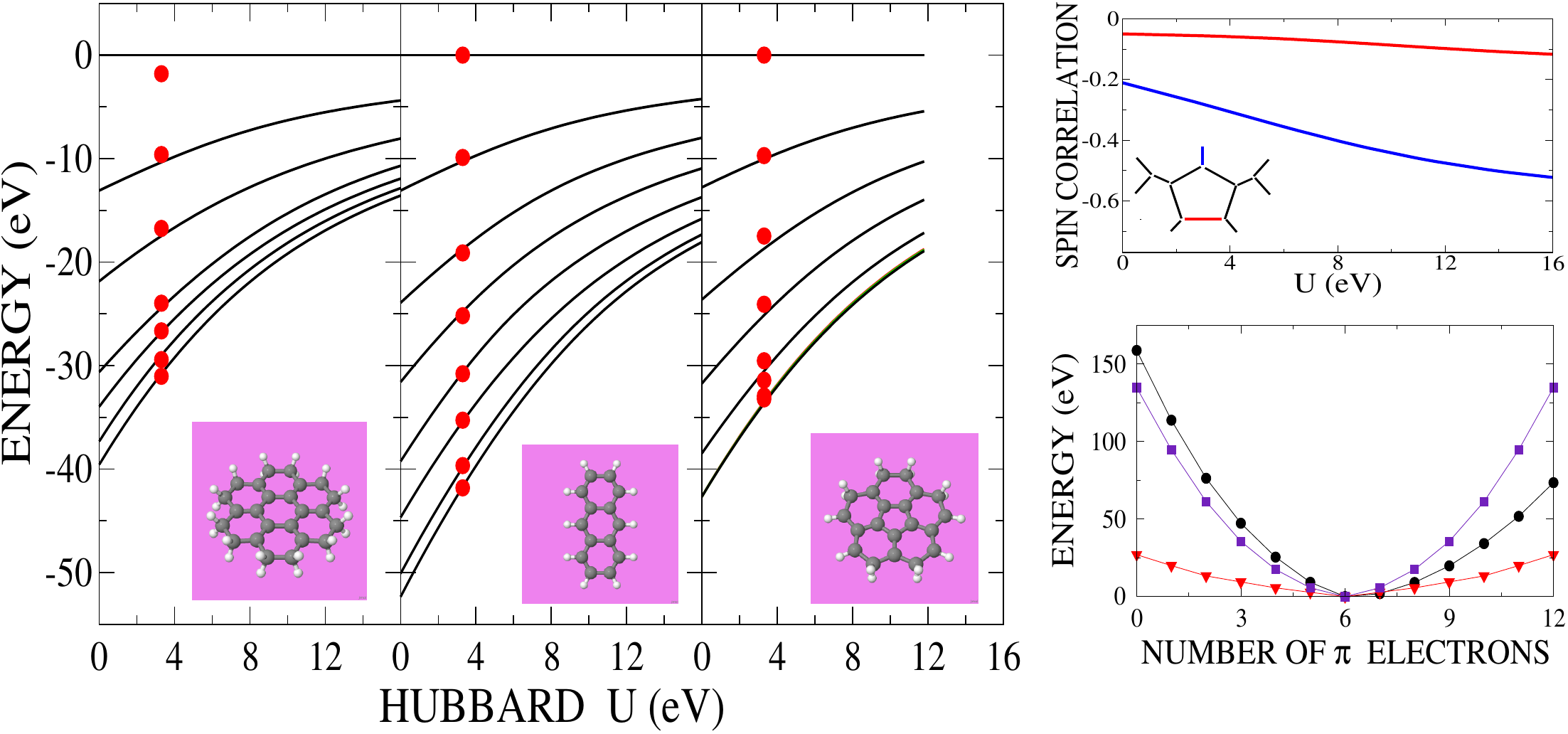}
\caption{(color online)
{\bf Left:} Total RB3LYP (circles) and Hubbard model (continuous curves)
energies of spin S states versus the on-site Coulomb repulsion U
for dodecahydrogenated coronene C$_{24}$H$_{24}$
(each peripheral carbon atom saturated with an additional hydrogen),
anthracene and 6H-corannulene C$_{20}$H$_{16}$ (Fig. 1C).
The sequence from lowest to highest energy is S=0,1, ... for
anthracene and C$_{24}$H$_{24}$ and S=2,1,0,3,4, ... for
C$_{20}$H$_{16}$ (in this molecule the Hubbard model
gives an almost twofold degenerate ground state).
Energies are  referred to the corresponding state of
maximum spin, except for C$_{24}$H$_{24}$ that were downward shifted
by 1.8 eV to improve the overall fit
(the state of S=6 is largely participated
by H orbitals turning invalid the Hubbard model).
Molecular geometries were only optimized for the ground state and
taken unchanged for the calculation of excited spin states.
Transfer integral was taken equal to -2.71 eV, and, as the results indicate,
U = 3.3 eV nicely reproduces the
{\it ab initio} energies. {\bf Right up:} Spin-spin
correlations in C$_{20}$H$_{16}$ calculated by means of
the Hubbard model (the skeleton of C atoms
is shown as an inset) on blue pair
(two atoms placed on the symmetry axis)
and red pair (frustrated horizontal bond).
{\bf Right down:} {\it Ab initio} ground state
energies (circles) of the charged states of a benzene molecule
that is not allowed to relax.
Energy differences are plotted relative to the neutral case.
Results obtained by means of Hubbard (triangles) and
Pariser-Parr-Pople (squares) models are also shown.}
\label{fig:energies-abin}
\end{figure*}

\section{Model Hamiltonians}
Let us critically examine the applicability of Lieb's theorem as the predicting
tool of the multiplicity of the ground state of hydrogenated PAHs.
The underlying Hubbard model ignores that:
(i) transfer integrals in any realistic system are not limited to
nearest neighbors sites,
(ii) $\sigma$--orbitals appear around the HOMO-LUMO gap in the same
energy interval as $\pi$--orbitals,
(iii) interaction among electrons is not limited to
on-site Coulomb repulsion. In our opinion,
the success of a theorem or rule based on the simplest interacting
model comes from its actual capability of describing the correct
antiferromagnetic spin-spin correlations between nearest $\pi$ electrons.
Strong correlation is the basis for the basic correctness of a
simplified image in which up and down spins alternate \cite{note}.

Even if the spin multiplicity of the ground state is predicted either by
Hund rule or Lieb's theorem, a deeper understanding of underlying correlations
calls for a complete numerical solution of simple interacting models.
We have analyzed both Pariser-Parr-Pople (PPP) model
Hamiltonian \cite{PP53,Po53}
and the local version of Hubbard Hamiltonian \cite{Hu63}, which actually is a
particular case of the former. The PPP Hamiltonian contains
a non-interacting part $\hat H_{0}$ and a term that incorporates the
electron-electron interactions $\hat H_{I}$:

\begin{equation}
{\hat H}  = {\hat H_{0}} + {\hat H_{I}}
\label{eq:H}
\end{equation}

\noindent
The non-interacting term is written as,

\begin{equation}
{\hat H_{0}}  = \epsilon_0 \sum_{i=1,N;\sigma}
c^{\dagger}_{i\sigma} c_{i\sigma} +
t \sum_{<ij> ; \sigma}c^{\dagger}_{i\sigma} c_{j\sigma}
\label{eq:H_{0}}
\end{equation}

\noindent
where the operator $c^{\dagger}_{i\sigma}$ creates an electron
at site $i$ with spin $\sigma$, $\epsilon_0$ is the energy of carbon
$\pi$--orbital,
and $t$ is the hopping between nearest neighbor pairs (kinetic energy).
$N$ is the number of unsaturated C atoms.
The interacting part is in turn given by,

\begin{equation}
{\hat H_{I}}  =
U\sum_{i=1,N;\sigma}n_{i\uparrow}n_{i\downarrow}+{\frac{1}{2}}
\sum_{i\neq j;\sigma,\sigma'}V_{|i-j|} n_{i\sigma} n_{j\sigma'}
\label{eq:H_{I}}
\end{equation}

\noindent
where $U$ is the on-site Coulomb repulsion and $V_{|i-j|}$ is
the inter-site Coulomb repulsion, while the density operator is

\begin{equation}
n_{i\sigma}= c^{\dagger}_{i\sigma} c_{i\sigma}\;.
\label{eq:density}
\end{equation}

\noindent
This Hamiltonian reduces to the Hubbard model for $V_{|i-j|}$=0

We start discussing the fitting of spin state energies by means
of Hubbard Hamiltonian. Lanczos algorithm in the whole Hilbert
occupation space is used to get numerically exact
many-body states \cite{lanczos}.
Coulomb on-site repulsion has been adjusted to describe
spin excitations of some PAHs.
Because the interacting model cannot be solved exactly for 6H-coronene
(18 orbitals or equivalently sites lead to a Hilbert occupation space
of dimension equal to $4^{18}$ which is beyond actual
computational facilities), the case of a coronene molecule
with all peripheral C atoms saturated by additional H has been considered.
This leaves a molecule with only 12 $\pi$ orbitals,
a cluster size that can be easily handled by means of Lanczos algorithm.
Also benzene (6 sites), anthracene (14 sites) and
6H-corannulene (14 sites) have been fitted.
Calculations were carried out by taking the hopping
integral commonly used to describe graphene sheets, $t = - 2.71$ eV,
and varying the on-site repulsion $U$. The results depicted in the
left panel of Fig. 4
indicate that spin states of these molecules can be
reasonably fitted with $U = 3.3$ eV
(benzene, for which the fitting is as good as for anthracene,
is not shown for the sake of clarity).
Albeit noticeable deviations occur in the
three lowest lying states of 6H-corannulene,
the state ordering is the correct one. We have checked that
the failure to correctly separate the lower excitations of 6H-corannulene
is not exclusive of the simplest interacting model: a PPP calculation
using Ohno's interpolation scheme \cite{Oh64} shows a similar weakness.
Let us remark once more that, despite
the rather small $U$ ($U / |t| = 1.27$) resulting from the fittings
shown in Fig. 4, anti-ferromagnetic correlations in these molecules,
as calculated by means of the interacting model, are significant.
Particularly attracting is the case of corannulene
for which there is a bond at which the spin-spin correlation is
significantly smaller than at other bonds of the molecule
(top right panel of Fig. 4).
Interestingly enough, placing frustration (two adjacent $\pi$ orbitals of
the same sublattice as shown in Fig. 1C) at that bond
gives a difference between carbon atoms in the two sublattices of four atoms,
which, using Lieb's formula, predicts a ground state of total spin 2,
in agreement with our numerical results.
Summarizing, $U$ = 3.3 eV works satisfactorily describing the spin
states of aromatic molecules and, in particular, the
multiplicity of the ground state.

The same simple Hubbard model fails, however, in describing the
charged states of these systems as results for benzene show
(bottom right panel of Fig. 2).
Lanczos results for the interacting Hamiltonian are
compared with B3LYP results for a charged benzene ideally restricted to
a fixed geometric structure. Actual energy differences are much higher
than those predicted by the model.
Our results do also illustrate the lack of electron-hole symmetry
that characterizes any realistic self-consistent field calculation
as opposed to Lieb's model. The PPP model with $t = - 2.71$ eV and
values for the Coulomb repulsion integrals from Ref. \onlinecite{PP53},
although greatly improves the fitting, still gives a symmetric
curve by the well-known pairing between occupied and unoccupied molecular
orbitals. A full fitting of charge and spin states will surely require
including a larger number of parameters \cite{OV03,CL07}.

\section{Concluding Remarks}
We have proposed a new
route to produce magnetic organic molecules that consists of
hydrogenating PAHs. In the case of alternant PAHs the
spin multiplicities of the molecules ground states
agree with Lieb's prediction,
even though {\it ab initio} Hamiltonians may significantly differ
from Hubbard's  model. A probably related result is that {\it ab initio}
energies of the spin states of these 
molecules can be very satisfactorily fitted by means of the simplest
version of the Hubbard model. 
It seems that the molecule topology is enough support for the main result.
Energies of charged molecules, instead, cannot be described by the simple,
most popular, interacting models,
suggesting a critical examination of their use in, for instance, graphene.
Results for total spin densities in molecules having a magnetic ground state
clearly show that 
the spin is localized in only one of the two sublattices.
On the other hand, {\it ab initio} and model Hamiltonian calculations
for hydrogenated corannulene, place the frustrated bond at the same location.
This produces an unbalance in the molecule which, using Lieb's formula,
gives S=2 for the ground state of the molecule,
in agreement with the numerical results.
It is also worth noting that in a recent study \cite{VS09} we have
shown that dehydrogenation may also produce magnetic molecules.
Although dehydrogenation is a highly unlikely process,
dehydrogenated PAHs have been intensively investigated by
astrophysicists \cite{LS03} who believe they to form part of
interstellar matter.
Although the results presented here are encouraging,
there is still a long way to go:
finding procedures to synthesize these hydrogenated PAHs and
crystallize them into solids that may eventually show magnetic properties.

\begin{acknowledgments}
The authors are grateful to J. Feliu, M. Yus and A. Guijarro for useful
suggestions and remarks.
Financial support by the Spanish MCYT (grants FIS200402356, MAT2005-07369-C03-01
and NAN2004-09183-C10-08)
and the Universidad de Alicante is gratefully acknowledged.
GC is thankful to the Spanish MCYT for a Ram\'on y Cajal grant.
\end{acknowledgments}


\begin{thebibliography}{99}
\bibitem{UHF} Remember that Unrestricted-Hartree-Fock calculations
do not provide wavefunctions with a defined total spin (only one
component is a good quantum number).
\end{thebibliography}
\end{document}